\documentclass[a4paper]{jpconf}
\usepackage{amssymb}
\usepackage{amsmath}
\usepackage[dvipdfm]{graphicx}

\begin{document}
\title{The cross-correlation search for a hot spot of gravitational waves : Numerical study for point spread function }

\author{Yuta Okada$^1$, Nobuyuki Kanda$^{1*}$, Sanjeev Dhurandhar$^2$, Hideyuki Tagoshi$^3$ and Hirotaka Takahashi$^{4,5}$}

\address{$^1$ Graduate School of Science, Osaka City University,  Sumiyoshi-ku, Osaka 558-8585, Japan}
\address{$^2$ Inter-University Centre for Astronomy and Astrophysics, Post Bag 4, Ganeshkhind, Pune 411007, India}
\address{$^3$ Department of Earth and Space Science, Graduate School of Science, Osaka University, Toyonaka, Osaka 560-0043, Japan} 
\address{$^4$ Department of Humanities, Yamanashi Eiwa College, 888, Yokone, Kofu, Yamanashi 400-8555, Japan}
\address{$^5$ Earthquake Research Institute, University of Tokyo, Bunkyo-Ku, Tokyo 113-0032, Japan}

\address{$^*$ Paper presented at the conference by N. Kanda}

\ead{kanda@sci.osaka-cu.ac.jp}

\begin{abstract}
The cross-correlation search for gravitational wave, which is known as 'radiometry', has been previously applied to map of the gravitational wave stochastic background in the sky and also to target on gravitational wave from rotating neutron stars/pulsars. 
We consider the Virgo cluster where may be appear as `hot spot' spanning few pixels in the sky in radiometry analysis.
Our results show that sufficient signal to noise ratio can be accumulated with integration times of the order of a year.
We also construct numerical simulation of radiometry analysis, assuming current constructing/upgrading ground-based detectors.
Point spread function of the injected sources are confirmed by numerical test. Typical resolution of radiometry analysis is a few square degree which corresponds to several thousand pixels of sky mapping.
\end{abstract}

\section{Introduction}
Current constructing and upgrading ground-based gravitational wave detectors may reach the strain sensitivity in order of $10^{-24} {\rm [1/\sqrt{Hz}]}$ in near future. These detector sites are dislocated, and make possible to have a independent observation of gravitational waves.
Since the dislocated detectors may not have a correlation in noise component but have a correlation in gravitational waveforms, cross-correlation product by two or more detectors make possible to search for unknown gravitational waveforms. This is useful property in burst gravitational wave search for short duration, or in stochastic gravitational wave search for long duration. 
Distance between detectors also make possible triangulation for signal measurement. Time delay between detectors suggest the incident direction of the gravitational wave. It is used in short duration signal analysis. On the other hand, an employment of the time delay term in long duration signal search or stochastic gravitational wave which come from any directions is not simple.

However, we can extract the signal strength if we resolve Earth's rotation and convoluted signal from anywhere in the sky.
Using multiple detectors and long time integration, the deconvolution will be possible.
The directional cross-correlation search for gravitational wave is proposed as 'radiometry'.

Radiometry and any other triangulation analysis require long distance among detectors, and variation of arrangement (zenith direction, and
azimuthal rotation of interferometers) make possible to cover the sky and resolve the polarization. Recent starting of the construction of LCGT\cite{LCGT} in addition to previously constructed advanced detector sites of LIGO\cite{LIGO} and Virgo\cite{VIRGO} encourage the development of radiometry analysis.

\subsection{Radiometry Filter}

\begin{figure}[htb]
\begin{center}
\includegraphics[width=9cm]{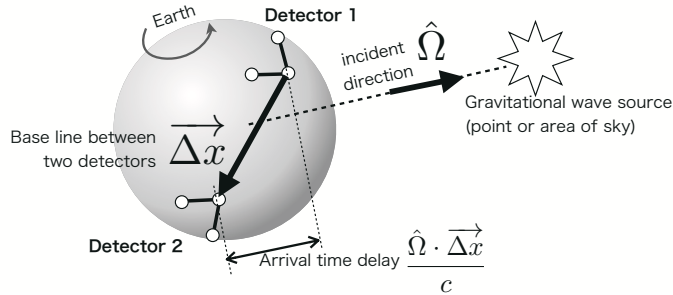}
\caption{Ideal setup of a radiometry}
\label{fig:IdeaOverview}
\end{center}
\end{figure}

Gravitational wave radiometry is a kind of stochastic gravitational wave search, but it is a directional analysis. 
Radiometry analysis assumes that gravitational wave come from particular direction as a point like source or a hot spot of sky.
The radiometry filter uses at least two detector observation data, and do not require the waveform.
Gravitational wave coming from particular direction will have a time delay of arrival at two detectors shown in Fig.\ref{fig:IdeaOverview}.
Since gravitational wave signals on both detectors are coherent in same polarization component, we can extract gravitational wave signals
in the cross-correlation product from two detectors observation data with appropriate time delay. Moreover, 
signal-to-noise ratio for gravitational wave will be emphasized with long integration time, because the two detector noises are independent.
Changing the assumption of source direction step by step, we can search gravitational wave `sky map' by radiometry analysis.

Radiometry analysis is given and studied in previous works\cite{mitra}\cite{bose}. We explain the radiometry filter briefly here.
The source direction or particular direction of search can be represent as unit vector $\hat{\Omega}$. Two detectors are noted as 1 and 2,
and a base-line from detector 1 to detector 2 can be given as vector $\overrightarrow{\Delta x}$. 
An arrival time delay for gravitational wave between two detectors is $\displaystyle \frac{\hat{\Omega}\cdot\overrightarrow{\Delta x}}{c}$, where $c$ is a speed of light.
Detector output from each detector $\displaystyle s_{1}(t), s_{2}(t\ -\ \frac{\hat{\Omega}\cdot\overrightarrow{\Delta x}}{c})$ in time domain have same gravitational wave timing. 
In Fourier domain, this time delay represent as a phase difference of two detectors output $\tilde{s}_{1}(f), \tilde{s}_{2}(f)$ in Fourier transform of $s_1(t)$ and $s_2(t)$. Therefore, we can define the radiometry filter as following. A directed overlap reduction function $\gamma(f,\hat{\Omega})$ can be defined to correct phase difference as 
\begin{equation}
\gamma(f,\hat{\Omega})=\sum_{A=+,\times}\!\!\!F_1^A F_2^A e^{i 2\pi f \frac{\hat{\Omega}\cdot\overrightarrow{\Delta x}}{c}}\ ,
\label{eq:gamma}
\end{equation}
where $F_{1,2}^{A}$ are antenna pattern\cite{antennaPat} of each detector 1 or 2 for each two polarization $A=+$ or $\times$ of gravitational wave.
To make optimal filtering, we employ two detectors noise power spectrum $P_{1,2}(f)$ respectively, and assume that gravitational wave spectrum as $H(f)$. Radiometry filter can defined as
\begin{equation}
Q(f,\hat{\Omega})=\lambda\frac{\gamma^*(f,\hat{\Omega})H(f)}{P_1(f)P_2(f)}\ ,
\label{eq:Q}
\end{equation}
and filter output is
\begin{equation}
\Delta S(\hat{\Omega})=\int_{-\infty}^{\infty}df\ \tilde{s}_1^*(f) \tilde{s}_2(f)Q(f,\hat{\Omega}).
\end{equation}

The base-line $\overrightarrow{\Delta x}$ will change relatively in celestial coordinate according to Earth's rotation in the case of ground-based detectors. Therefore, we calculate $\Delta S$ at short time slice (`chunk') during $t - \Delta t/2$ and $t + \Delta t/2$. Fourier transforms of detector signals can be represented using short Fourier transform of $\tilde{s}_{1,2}^*(t;\!f)$, and $Q(f)$ can be denoted as $Q(t,f)$.
\begin{equation}
\Delta S(t,\hat{\Omega})=\int_{-\infty}^{\infty} df\ \tilde{s}_1^*(t;\!f) \tilde{s}_2(t;\!f) Q(t,\!f,\!\hat{\Omega}).
\label{eq:radiometryFilter}
\end{equation}

Processing many chunks for long duration observation data, we can get the statistics of $\Delta S$.
Since the gravitational wave component will appear in real part of $\Delta S$, we employ a mean $\mu(\hat{\Omega})$
\begin{equation}
\mu(\hat{\Omega})=\langle\Re[\Delta S]\rangle\ ,
\label{eq:mu}
\end{equation}
 and a standard deviation $\sigma(\hat{\Omega})$ for a real part of $\Delta S$.
Signal-to-noise ratio of particular direction can be given as
\begin{equation}
\rho(\hat{\Omega})=\mu(\hat{\Omega})/\sigma(\hat{\Omega}).
\label{eq:sn}
\end{equation}

\section{Hot Spot of Gravitational Wave Radiometry}

Radiometry analysis might find a hot spot in sky map. Hot spot may be formed by concentration of unresolved huge number of astronomical sources.
Here is one of the promising targets the Virgo cluster which could contain $\sim 10^{11}$ neutron stars. 
Our results show that sufficient signal to noise can be accumulated with integration times of the order of a year. We published this study in the reference \cite{OurRadiometryPaper}. If an average ellipticity of neutron stars are $10^{-6}$, Virgo cluster might be appear as a hot spot with observation of order of year, using 2nd generation ground-based detectors.
The numerical estimation of expected GW spectrum and signal-to-noise ratio of Virgo cluster hot spot is explained in this reference in detail.

Here we remark the average detector response for the sky.
In eq.(\ref{eq:gamma}), magnitude of two detector's combined detector response $\Gamma(\hat{\Omega}, t)$ is give as
\begin{equation}
\Gamma(\hat{\Omega}, t) = F_{+1}(\hat{\Omega}, t) F_{+2}(\hat{\Omega}, t) + F_{\times 1}(\hat{\Omega}, t) F_{\times 2}(\hat{\Omega}, t).
\end{equation}

\begin{figure}[htbp]
\begin{center}
\includegraphics[width=8cm]{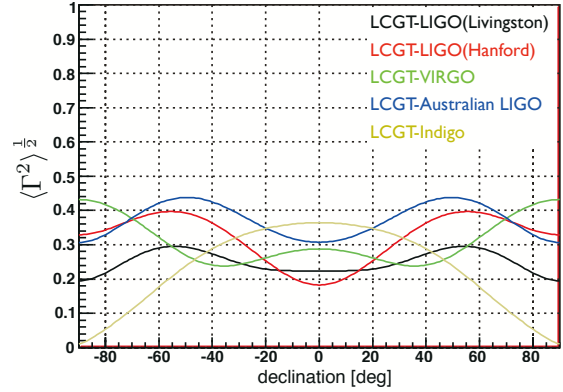}
\caption{Average of detector response $\Gamma^2$ as a function of celestial declination angle.}
\label{fig:gammaSQ}
\end{center}
\end{figure}

Since $\Gamma(\hat{\Omega}, t)$ will change according to Earth's rotation, we have to use average during one day :
\begin{equation}
\langle\Gamma^2\rangle_{1\mathrm{day}}(\hat{\Omega})=\frac{1}{T_{1\mathrm{day}}}\int_0^{1\mathrm{day}}\hspace{-5pt}\Gamma^2(\hat{\Omega},t)\,dt.
\label{eq:CapitalGamma}
\end{equation}
$\langle\Gamma^2\rangle$ will not depend on a right ascension. Fig.\ref{fig:gammaSQ} displays square root of $\langle\Gamma^2\rangle$. 
For Virgo cluster whose center is about at declination $12^{\circ}4'59''$, LCGT and India site (IndIGO\cite{IndIGo}) is a good combination. Variation of combination will cover the whole sky in radiometry analysis.

\section{Numerical Simulation : Source Injection Test}

To confirme the behavior of a hot spot in the radiometry sky map, we construct numerical test of radiometry filter.
It consist of the following steps.

\begin{figure}[bhtp]
\begin{minipage}[t]{0.48\textwidth}
\begin{center}
   \includegraphics[width=1.0\textwidth, trim=55 20 5 50, clip]{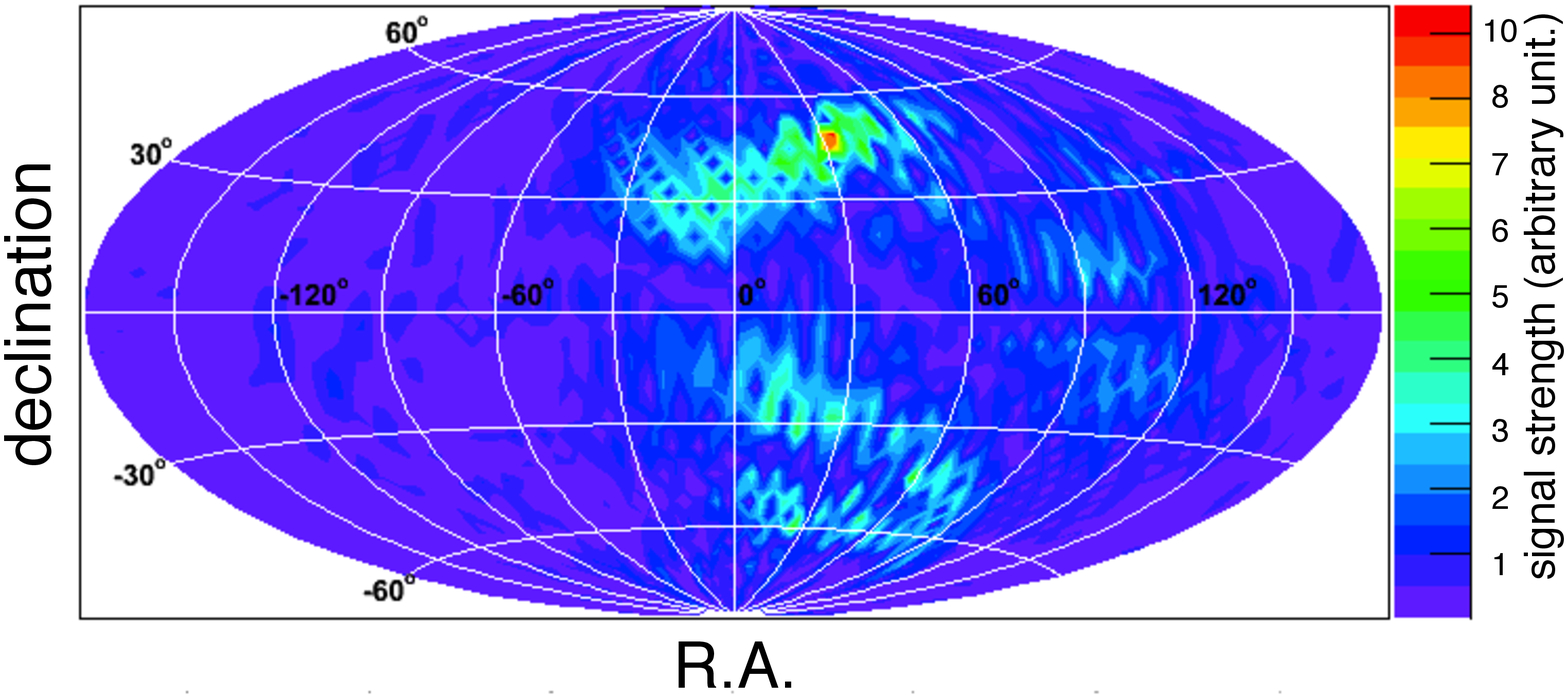}
   \caption{Example of sky map of radiometry output. This map is displayed in celestial coordinate.}
   \label{fig:mapExample}
\end{center}
\end{minipage}
\hspace{0.02\textwidth}
\begin{minipage}[t]{0.48\textwidth}
\begin{center}
   \includegraphics[width=1.0\textwidth]{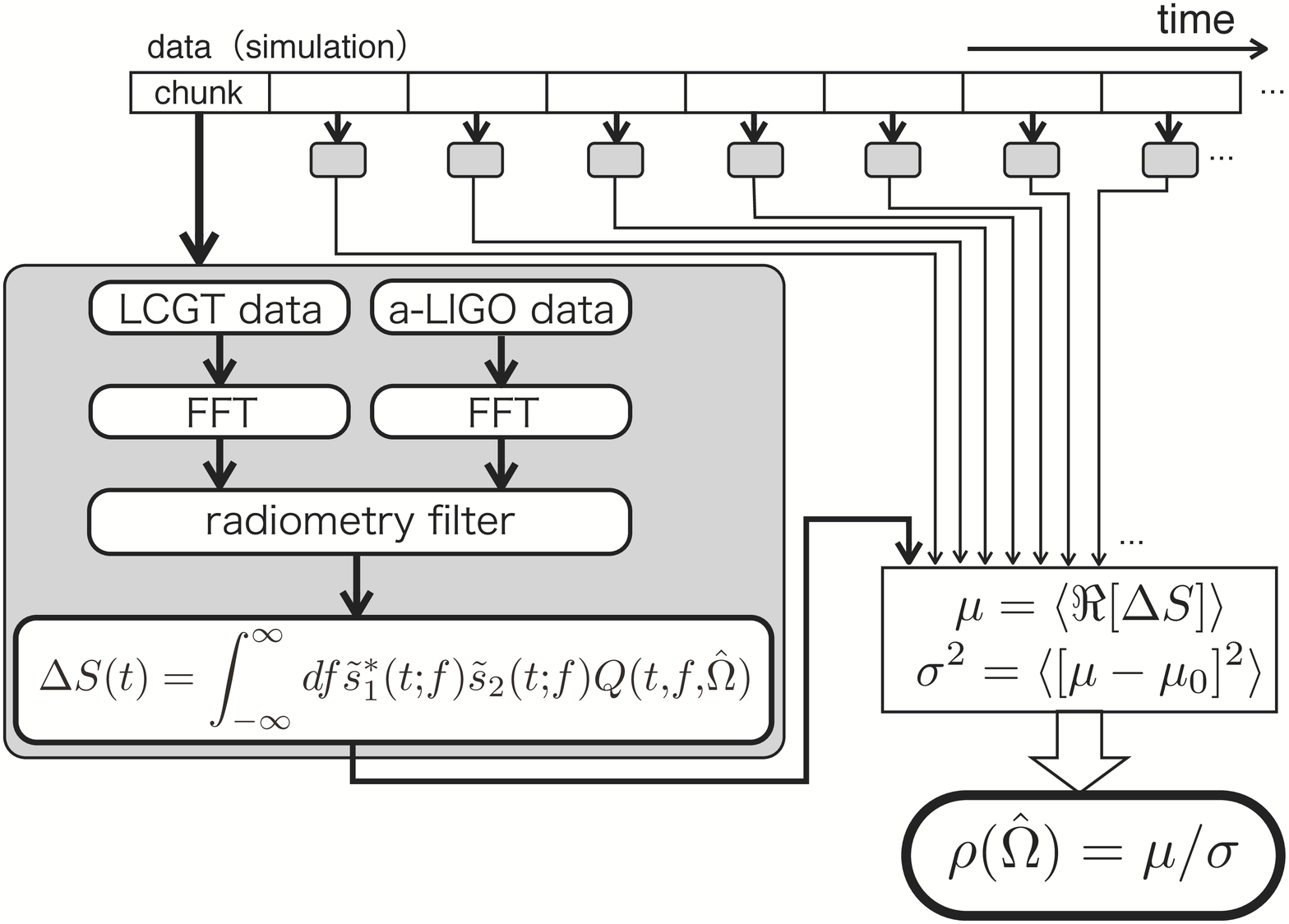}
   \caption{Data flow of numerical test of radiometry filter}
   \label{fig:DataFlow}
\end{center}
\end{minipage}
\end{figure}

\begin{description}
\item[ 1) set short duration of data stream along time axis as {\it `chunks'} : ]

In real observational data processing, this step and following steps 2) 3) 4) are corresponding to split time series data into short duration as {`chunks'}.
Following steps 2)--5) are processed for each chunk.

\item[ 2) generate gravitational wave detector noise : ] 

Assuming design noise spectrum \cite{spe_LCGT} \cite{spe_aLIGO} \cite{spe_aVirgo} \cite{spe_ET} of each detector and injecting random gaussian noise, we generate time series of noise $n(t)$. 
In this paper, time series data's sampling frequency is 4000 Hz, and chunk length is 65.536 sec.
In the study of point spread function in next subsection, we skip noise generation to use gravitational wave signal only.

\item[ 3) generate gravitational wave signal : ]

Gravitational wave signal at source is generated. We tested with sinusoidal wave, and random phase
 waveform which has power spectrum in certain frequency band, e.g. flat power spectrum in 300$\pm$30 Hz.

\item[ 4) inject gravitational wave signal into detector noises in time domain : ]\ 

Gravitational wave signal will incident detectors with Doppler effect, antenna patterns and time delay between two detectors.
Doppler effect is employed as a frequency modulus of gravitational wave on the detector according to the orbital motion and Earth's rotation.
For incident direction on detectors, antenna patterns $F_{+}, F_{\times}$ are calculated for the signal.
Also the arrival time difference of the signal on two (or more) detectors are given in this step.
These condition will change as time goes by according to Earth's rotation.

\item[ 5) calculate radiometry filter output $\Delta S (t,\hat{\Omega})$ : ]

We calculate radiometry filter output of this chunk in eq.(\ref{eq:radiometryFilter}) using two detectors
for the particular direction $\hat{\Omega}$ of a {\it `pixel'}. Changing $\hat{\Omega}$ step by step, sweeping all sky, we calculate $\Delta S (t, \hat{\Omega})$ for all pixels.

\item[ 6) repeat step 1) -- 5) for next chunk : ] 
We iterate steps above for all chunks.
\\

Finnaly,

\item[ 7) sum up $\Delta S$ and calculate signal-to-noise ratio $\rho(\hat{\Omega})$ of each pixel : ]

We sum up each cunk's output $\Delta S (t,\hat{\Omega})$ with respect to each pixel. Then we get $\mu(\hat{\Omega}), \sigma(\hat{\Omega})$ and
calculate signal-to-noise ratio $\rho(\hat{\Omega})$ in eq.(\ref{eq:sn}). $\rho(\hat{\Omega})$ are displayed as pixel of sky map with colorized according to its magnitude as Fig.\ref{fig:mapExample}.

\end{description}

Fig.\ref{fig:DataFlow} displays a schematic data flow of our numerical test of radiometry filter of above steps. We tried possible combinations with current detectors, but Fig.\ref{fig:DataFlow} displays LCGT and advanced LIGO for example. For studying point spread behavior in following sections of this paper, we display also in the case of LCGT and advanced LIGO at Livingston.

\subsection{Point Spread Function}

How spread the point source on the sky is important to evaluate the feasibility of searches. It is called as {\it `point spread function'}.
Point spread function depends on gravitational wave frequency or spectrum, a position on the sky (right ascension, celestial declination), and for detector orientation on the Earth. In point spread function study in this subsection, we used noiseless detector output, but other filter setting is same to explanation above. Thus, the magnitude of the gravitational wave in this section is arbitrary scale. Test injection of gravitational wave signal has a flat spectrum around $f_0$ [Hz] as
 
\begin{eqnarray}
H(f)=
\left\{
\begin{array}{cl}
1&( 0.9 f_0 \le f\le 1.1 f_0)\\
0&(\mathrm{else})
\end{array}
\right.
\end{eqnarray}
for average spectrum. Injected signal for each chunk fluctuates as stochastic signal whose phase is random.

\begin{figure}[htbp]
\begin{center}
\includegraphics[width=15cm]{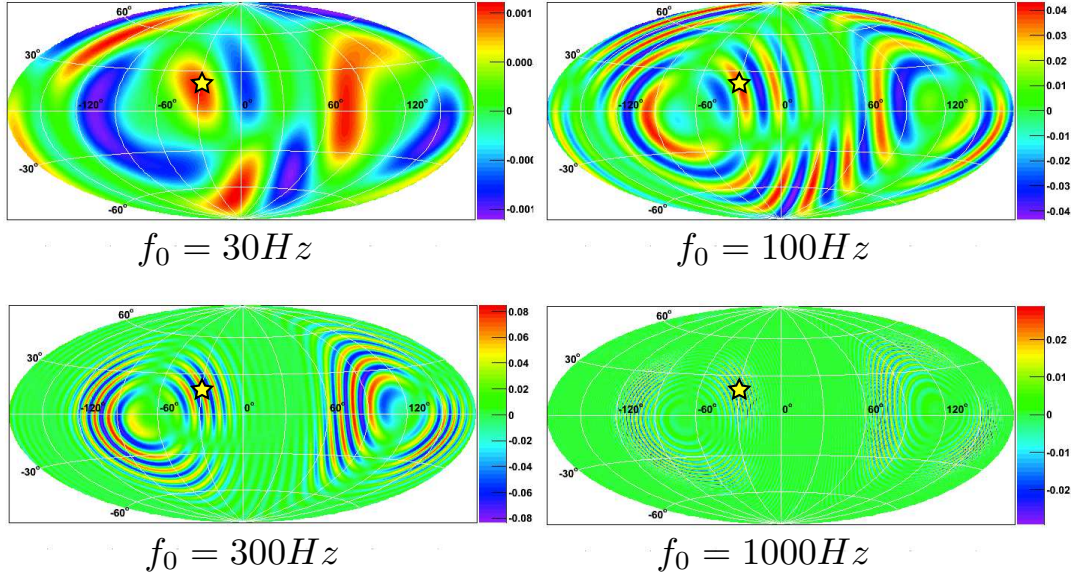}
\caption{Radiometry filter output of single chunk with various frequency signal. Star symbol shows true injection position in the sky. Gravitational wave has a flat spectrum around $f_0$ [Hz] with $\pm$ 10\% of $f_0$ [Hz] width; e.g. $f_0$=300Hz means from 270Hz to 330Hz.}
\label{fig:GlobalMap_Aliases}
\end{center}
\end{figure}

Fig.\ref{fig:GlobalMap_Aliases} shows how point spreading for some frequency samples in single chunk process. In single chunk, we can find of wide and multiple spreading image from the injection of point source, since the different time lag makes aliases. Spatial interval is wide for low frequency band, and narrow for higher frequency band. But in general, it is hard to identify the exact position of the point source in single chunk map, even if the gravitational wave is so strong.

\begin{figure}[htbp]
\begin{center}
\includegraphics[scale=0.48]{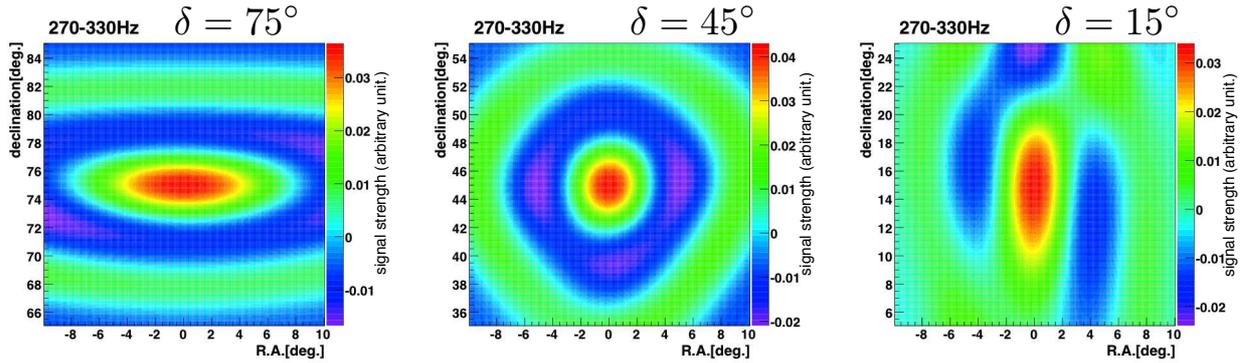}
\caption{Point spread image in radiometry map (closed up narrow region) with one day integration with variation of declination angle. Frequency $f_0$ of injected signal is 300 Hz.
Declination angles of injection point is $\delta = 15^{\circ}, 45^{\circ} {\rm and}\ 75^{\circ}$ respectively.}
\label{fig:spread_delta}
\end{center}
\end{figure}

With Earth's rotation, aliases will change its orientation since detector position will change relatively to the source. Therefore, during long integration time, aliases will be reduced and true position will grow up with stacking filter output coherently.
Fig.\ref{fig:spread_delta} displays radiometry map with one day integration for closed up narrow region around injection sources. 
The figure displays variation in case of declination angle of $\delta = 15^{\circ}, 45^{\circ} {\rm and}\ 75^{\circ}$.
The point spread function is not depend on right ascension with more than one day integration.
According to the combined detector response as eq.(\ref{eq:Q}) with LCGT and LIGO detectors, mid declination angle have a circular image of spreading. 
On the other hand, in low and high declination area, point spreads as ellipsoid.

\begin{figure}[htbp]
\begin{center}
\includegraphics[scale=0.48]{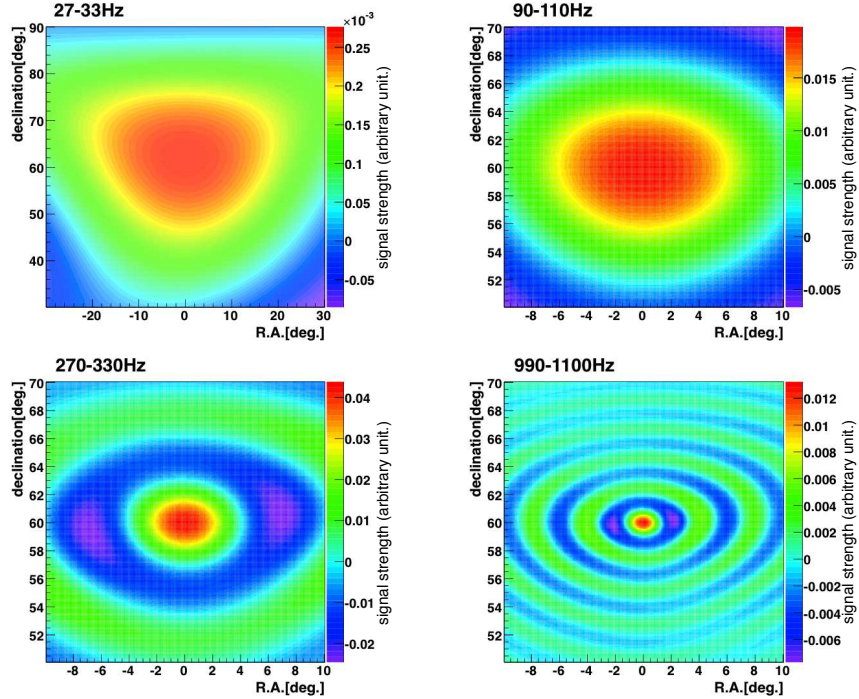}
\caption{Point spread image in radiometry map with one day integration with variation of gravitational wave frequency. Injection point is at $\delta = 60^{\circ}$. 
Central frequency of each figure is 30Hz, 100Hz, 300Hz and 1kHz respectively.}
\label{fig:spread_f}
\end{center}
\end{figure}

Spread size of the point source is strongly depend on the gravitational wave frequency. 
Its order is determined by a diffraction limit $\displaystyle \frac{\lambda}{|\overrightarrow{\Delta x}|}$, where wave length $\lambda$ of gravitational wave and distance between two detectors $|\overrightarrow{\Delta x}|$.
Fig.\ref{fig:spread_f} displays some examples of spread image of injection point at $\delta = 60^{\circ}$. 
The signal of 30 Hz spreads widely. 1kHz signal has aliases as concentric circle, but the true injected point appear as intense peak.

\begin{figure}[htbp]
\begin{center}
\includegraphics[width=12cm, trim=0 0 50 50, clip]{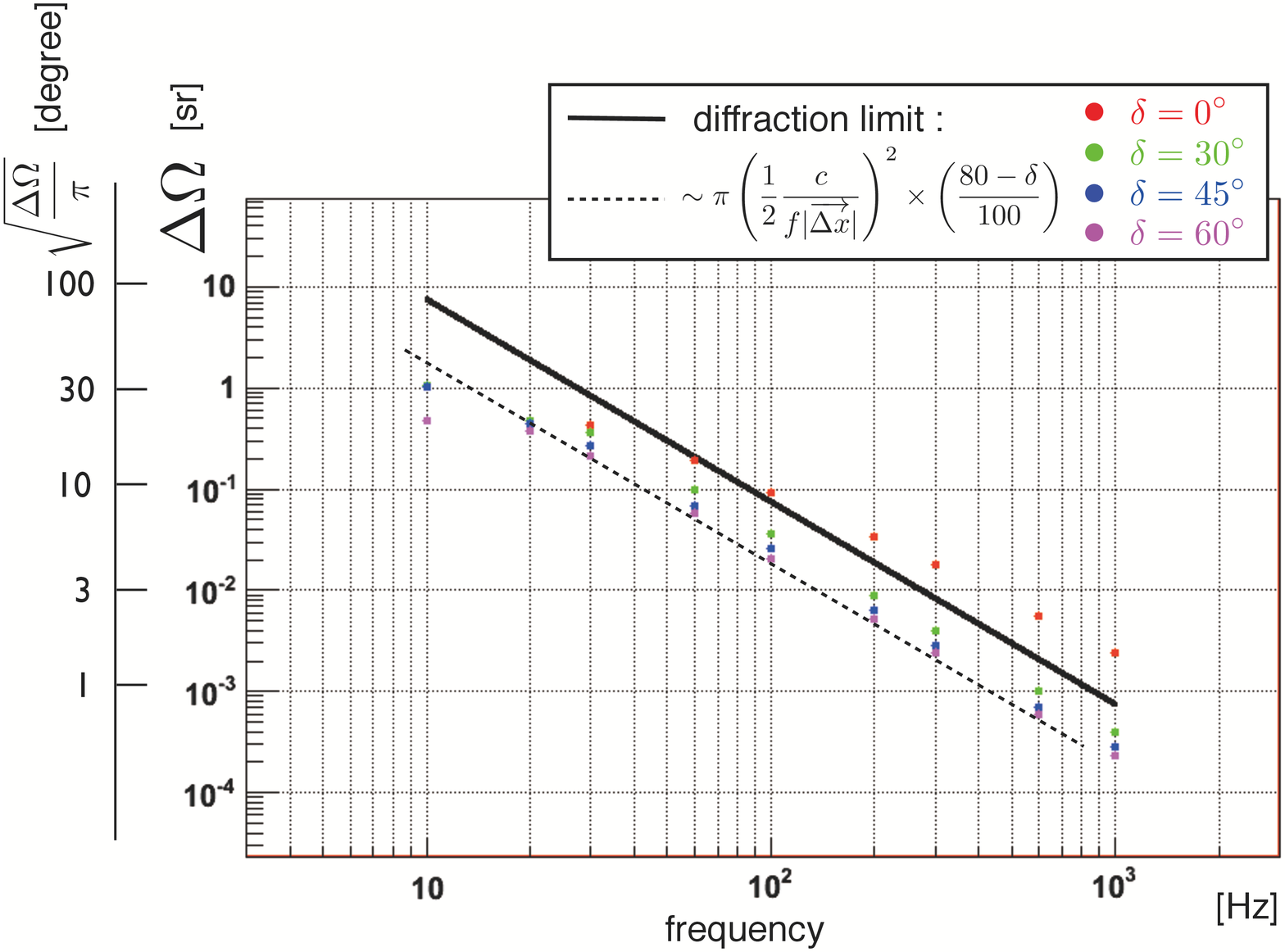}
\caption{Point spread size of injected gravitational waves. Points are injected gravitational wave signal with different central frequency$f_0$ and variation of declination angle $\delta$. $x$-axis is a gravitational wave frequency, and $y$-axis is a solid angle of point spreading. $y$-axis is also displays corresponding apparent diameter in case of circular image.}
\label{fig:diffraction_limit}
\end{center}
\end{figure}

To evaluate these spreading, we define the spread size using solid angle $\Delta \Omega$ [sr] as a solid angle of area which lager than half hight of the local peak.
Fig.\ref{fig:diffraction_limit} displays relation between $\Delta \Omega$ and the frequency of gravitational waves. Solid line is a $\displaystyle \frac{\lambda}{|\overrightarrow{\Delta x}|}$ and
dashed line is a empirical approximation for higher $\delta$ region :  
\begin{equation}
\sim\pi\left(\frac{1}{2}\frac{c}{f \ |\overrightarrow{\Delta x}|}\right)^2\times \left(\frac{80-\delta\ {\rm [deg.]}}{100}\right) {\rm [sr]}.
\end{equation}

In the Fig.\ref{fig:diffraction_limit}, we also display corresponding apparent diameter for $y$-axis in case of circular image.

\subsection{Optimal Pixel Size}

\begin{figure}[htbp]
\begin{center}
\includegraphics[width=12cm, trim=0 0 0 50, clip]{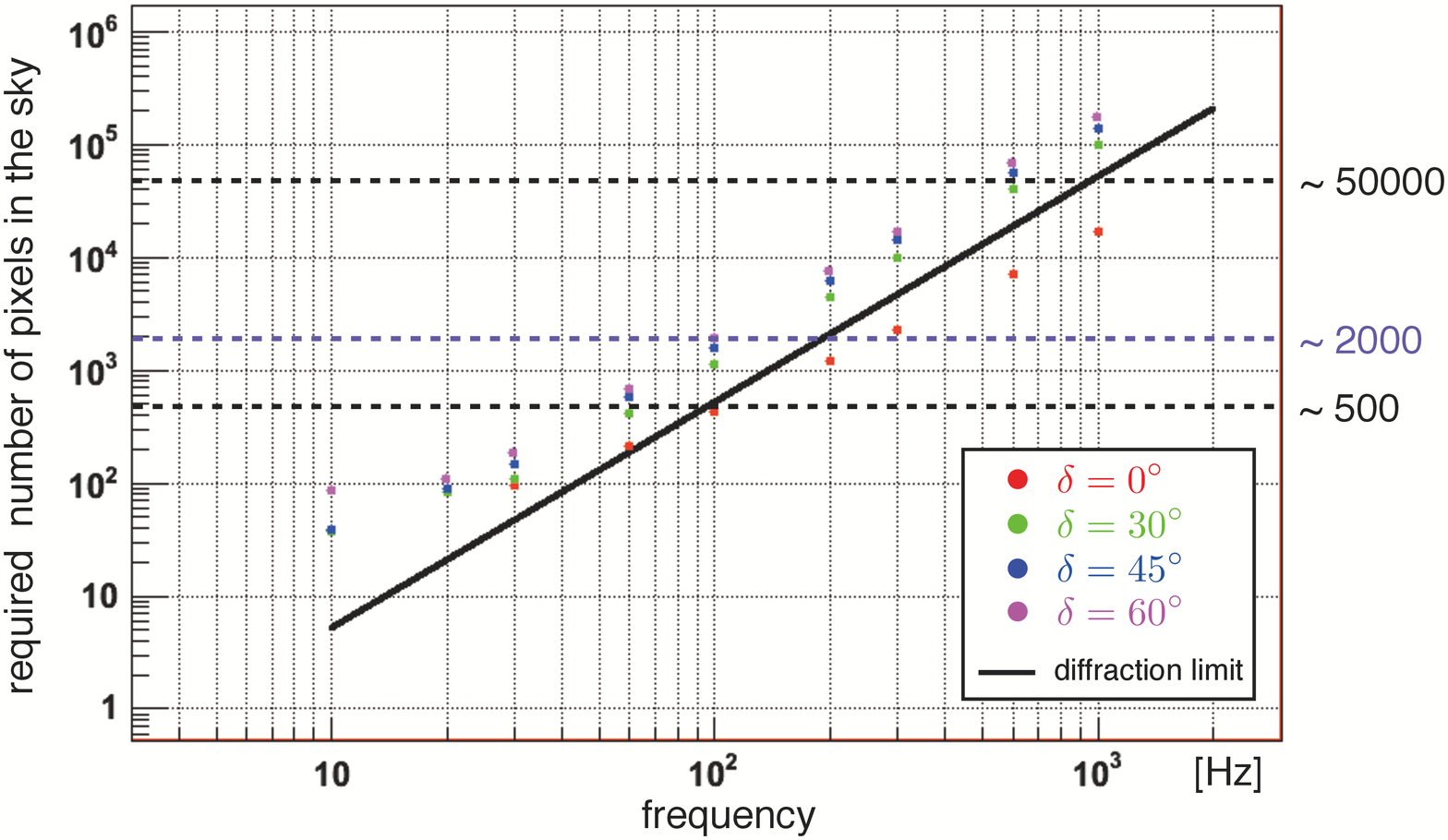}
\caption{Optimal number of pixels in the sky when we employ as point spread size (solid angle) as pixel size. Solid line is corresponding to the diffraction limit size.}
\label{fig:NumPix}
\end{center}
\end{figure}

Point spreading suggests on optimal size of pixels. It is possible to employ fine pixels as similar size of point spread function, because this is an angular resolution of radiometry analysis. However, there is no need to divide the celestial sphere so fine as one pixel size is far smaller than point spread function.
Fig.\ref{fig:NumPix} displays optimal number of pixels in the sky when we employ as point spread size (solid angle) as pixel size. In this study with LCGT and LIGO (Livingston) base line can employ up to several thousand pixels for detectors' good sensitivity frequency band around $100\sim 300$ Hz.

\section{Summary}

We studied radiometry analysis for stochastic gravitational waves from astronomical origin, which might have an anisotropy or local excess. 
Virgo cluster has a possibility to appear as a `hot spot' on the radiometry map of the sky with advanced and third generation gravitational wave detectors.
We also studied numerical simulation of radiometry filter, and displayed the spreading image for point like gravitational wave sources, with LCGT and advanced LIGO combination.
Also other combinations in global network detectors will be useful to emphasize the signal behavior.
We are planning to proceed the injection test of our numerical simulation studies for multiple sources and area spreading sources.

\ack
%S. Dhurandhar thanks S. Bose and S. Mitra for useful discussions on multiple baselines.
%We thank F. Takahara and S.J. Tanaka for useful discussions on the population of neutron stars. 
S. Dhurandhar 
acknowledges the DST and JSPS Indo-Japan international cooperative programme 
for scientists and engineers for supporting visits to Osaka City University,
Japan and Osaka University, Japan. 
H. Tagoshi, N. Kanda and H. Takahashi thank JSPS and DST under the same Indo-Japan programme for their visit to IUCAA, Pune, India.
N.Kanda's work was supported in part by a Monbu Kagakusho Grant-in-aid
for Scientific Research of Japan (No. 23540346).
H.Tagoshi's work was also supported in part by a Monbu Kagakusho Grant-in-aid
for Scientific Research of Japan (Nos. 20540271 and 23540309).
H.Takahashi's work was also supported in part by a Monbu Kagakusho Grant-in-aid
for Scientific Research of Japan (No. 23740207).

\end{document}